\documentclass[%preprint,
 superscriptaddress,
 floatfix,
 amsmath,amssymb,
twocolumn,
prl
]{revtex4-2}

\usepackage{graphicx}% Include figure files
\usepackage{dcolumn}% Align table columns on decimal point
\usepackage{bm}% bold math
\usepackage{amsmath}
\usepackage{hyperref}% add hypertext capabilities
\usepackage{lipsum}
\usepackage{physics}
\usepackage{tabularx}
\usepackage[caption = false]{subfig}
\usepackage{cleveref}
\renewcommand{\vec}[1]{\boldsymbol{\mathbf{#1}}}
\usepackage{slashed}
\usepackage{float}

 \DeclareGraphicsExtensions{.png,.pdf}
 
\begin{document}
\title{Novel spin-precession method for sensitive EDM searches}

\author{A.~Boeschoten}
\affiliation{Van Swinderen Institute for Particle Physics and Gravity, University of Groningen, The Netherlands}
\affiliation{Nikhef, National Institute for Subatomic Physics, Amsterdam, The Netherlands}
\author{V.R.~Marshall}
\affiliation{Van Swinderen Institute for Particle Physics and Gravity, University of Groningen, The Netherlands}
\affiliation{Nikhef, National Institute for Subatomic Physics, Amsterdam, The Netherlands}
\author{T.B.~Meijknecht}
\affiliation{Van Swinderen Institute for Particle Physics and Gravity, University of Groningen, The Netherlands}
\affiliation{Nikhef, National Institute for Subatomic Physics, Amsterdam, The Netherlands}
\author{A.~Touwen}
\affiliation{Van Swinderen Institute for Particle Physics and Gravity, University of Groningen, The Netherlands}
\affiliation{Nikhef, National Institute for Subatomic Physics, Amsterdam, The Netherlands}
\author{H.L. Bethlem}
\affiliation{Van Swinderen Institute for Particle Physics and Gravity, University of Groningen, The Netherlands}
\affiliation{Department of Physics and Astronomy, and LaserLaB, Vrije Universiteit Amsterdam, The Netherlands}
\author{A.~Borschevsky}
\affiliation{Van Swinderen Institute for Particle Physics and Gravity, University of Groningen, The Netherlands}
\affiliation{Nikhef, National Institute for Subatomic Physics, Amsterdam, The Netherlands}
\author{S.~Hoekstra}
\affiliation{Van Swinderen Institute for Particle Physics and Gravity, University of Groningen, The Netherlands}
\affiliation{Nikhef, National Institute for Subatomic Physics, Amsterdam, The Netherlands}
\author{J.W.F. van Hofslot}
\affiliation{Van Swinderen Institute for Particle Physics and Gravity, University of Groningen, The Netherlands}
\affiliation{Nikhef, National Institute for Subatomic Physics, Amsterdam, The Netherlands}
\author{K.~Jungmann}
\affiliation{Van Swinderen Institute for Particle Physics and Gravity, University of Groningen, The Netherlands}
\affiliation{Nikhef, National Institute for Subatomic Physics, Amsterdam, The Netherlands}
\author{M.C.~Mooij}
\affiliation{Nikhef, National Institute for Subatomic Physics, Amsterdam, The Netherlands}
\affiliation{Department of Physics and Astronomy, and LaserLaB, Vrije Universiteit Amsterdam, The Netherlands}
\author{R.G.E.~Timmermans}
\affiliation{Van Swinderen Institute for Particle Physics and Gravity, University of Groningen, The Netherlands}
\affiliation{Nikhef, National Institute for Subatomic Physics, Amsterdam, The Netherlands}
\author{W.~Ubachs}
\affiliation{Department of Physics and Astronomy, and LaserLaB, Vrije Universiteit Amsterdam, The Netherlands}
\author{L.~Willmann}
 \email{l.willmann@rug.nl}
\affiliation{Van Swinderen Institute for Particle Physics and Gravity, University of Groningen, The Netherlands}
\affiliation{Nikhef, National Institute for Subatomic Physics, Amsterdam, The Netherlands}

\collaboration{NL-\it{e}\rm{EDM} Collaboration}%\noaffiliation

\date{\today}% It is always \today, today,
             %  but any date may be explicitly specified

\begin{abstract}

{We demonstrate a spin-precession method to observe and analyze multi-level coherence between all hyperfine levels in the $X ^2\Sigma^+,N=0$ ground state of barium monofluoride ($^{138}$Ba$^{19}$F). The signal is sensitive to the state-preparation Rabi frequency and external electric and magnetic fields applied in searches for a permanent electric dipole moment (EDM). In the obtained interference spectrum, the electric field and Rabi frequency become observable simultaneously with the EDM. This method reduced systematic biases and the number of auxiliary measurements for such precision measurements.}
\end{abstract}
%\pacs{PACS need to be selected}

\maketitle

\textit{Introduction.}--- Searches for a permanent electric dipole moment (EDM) provide a sensitive test of discrete symmetries in the Standard Model \cite{ENGEL201321}. An EDM violates parity (P) and time-reversal (T) symmetry. Assuming the combined symmetry CPT is a symmetry of nature, where C is charge conjugation, T symmetry is equivalent to combined CP-symmetry. In composite systems, the effect of EDMs of elementary particles and other CP-violating interactions is greatly enhanced, which makes searches using atoms and molecules attractive \cite{sandars1975search}. Many experiments searching for atomic or molecular EDMs have been performed \cite{Regan2002,Hudson2011-2,Bishof2016,Graner2016,Cairncross2017,Andreev2018,https://doi.org/10.48550/arxiv.2212.11841, Zheng2022} or are proposed \cite{Vutha2018-2,Aggarwal2018,Verma2020, Grasdijk2021}. EDMs of paramagnetic molecules are in particular sensitive to the electron EDM $d_e$ and the scalar-pseudoscalar electron-nucleon interaction strength $C_{\rm{S}}$ \cite{Pospelov2022}. 
\begin{figure*}[t]
\includegraphics[width=.95\linewidth]{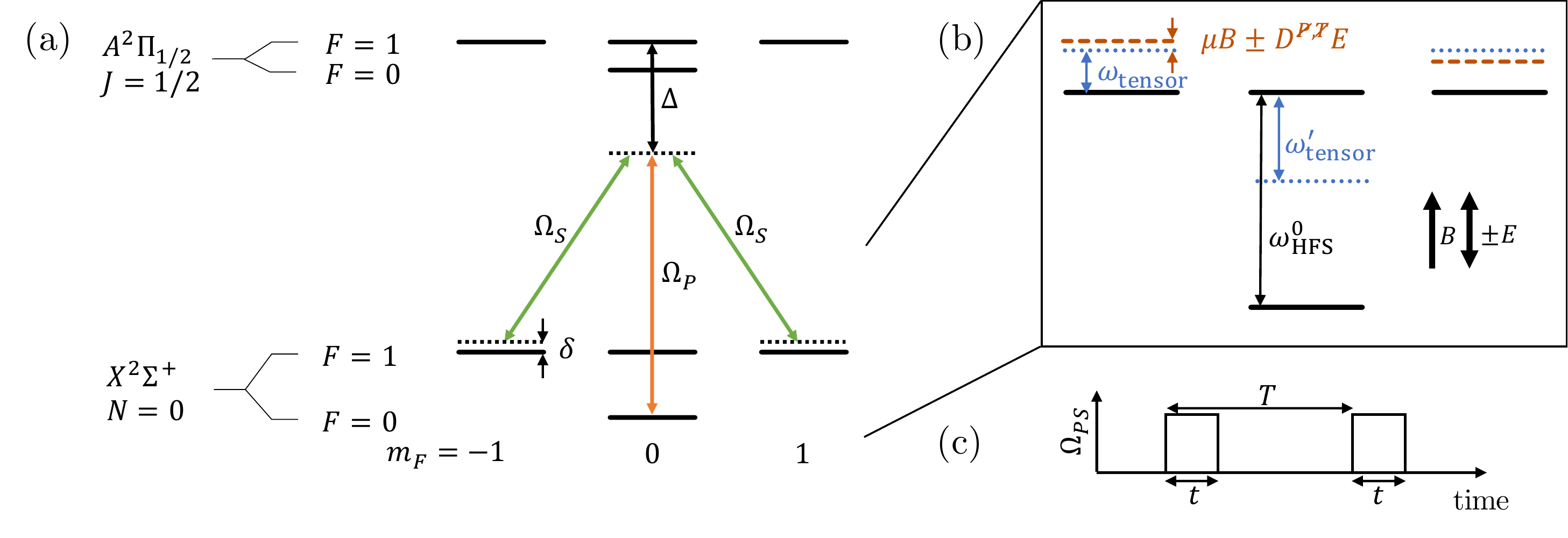}
\caption{(a) The $X^2\Sigma,v=0,N=0$ ground state and the electronically excited state $A^2\Pi_{1/2},v=0,J=1/2$ with hyperfine structure. The levels are coupled by two laser fields, labeled by $\Omega_S$ and $\Omega_P$, with orthogonal polarization, typical detuning $\Delta=1$~GHz from the $X^2\Sigma - A^2\Pi_{1/2}$ resonance, and two-photon detuning $\delta=\omega_{PS}-\omega_{\rm{HFS}}(E)$ of several kHz from two-photon resonance, where $\omega_{\rm{HFS}}(E)=\omega_{\rm{HFS}}^0+\omega_{\text{tensor}}(E)$. (b) The $X^2\Sigma,v=0,N=0$ sublevels of the ground state in electric and magnetic fields. The hyperfine splitting in absence of external fields $\omega_{\text{HFS}}^0$ is around $66~\text{MHz}$ and the tensor Stark shift for the $m_F=\pm1$ levels, $\omega_{\text{tensor}}(E)$, is around $10~\text{kHz}$. The tensor Stark shift for the $m_F=0$ level, $\omega'_\text{tensor}$, is twice that of $\omega_{\text{tensor}}$ with opposite sign. The eigenstates and energies in external fields are determined by diagonalization of an effective molecular Hamiltonian (e.g. \cite{BrownCarringtonBook2003}). (c) The timing sequence of the laser-light pulses with Rabi frequency $\Omega_{PS}$, where typical pulse lengths are $t = 80~\mu$s and the pulse separation period is $T=1$~ms. Energy levels and timings are not to scale.} 
\label{fig:expdescription}
\end{figure*}
The induced P,T-violating dipole moment of a molecule $\vec{D}^{\slashed{P},\slashed{T}}=D^{\slashed{P},\slashed{T}}\vec{F}/\hbar$, which is along its angular momentum $\vec{F}$, manifests as an extra splitting of magnetic sublevels in an external electric field ($\vec{E}$), and can be disentangled from the Zeeman shift, the interaction of the magnetic dipole moment $\vec{\mu}=\mu\vec{F}/\hbar$, with an external magnetic field ($\vec{B}$) due to the symmetries P and T. The Hamiltonian is $H = - \vec{\mu}\cdot\vec{B}-\vec{D}^{\slashed{P},\slashed{T}} \cdot\vec{E}$.
% \begin{equation}
%     H = - \vec{\mu}\cdot\vec{B}-\vec{D}^{\slashed{P},\slashed{T}} \cdot\vec{E}.
% \end{equation}
In EDM searches, typically electric and magnetic fields are applied that are either parallel or antiparallel with respect to each other. The different symmetries of $\vec{B}$ and $\vec{E}$ yield an energy dependence on the relative orientation of the fields for any non-zero value of $D^{\slashed{P},\slashed{T}}$. The aim of an experimental search for EDMs is an accurate determination of the energy splitting between selected Zeeman sublevels. The P,T-odd EDM contribution, interpreted in terms of $d_e$ and $C_{\rm{S}}$, is $D^{\slashed{P},\slashed{T}}E=-\frac{1}{2}(d_e W_d+C_{\rm{S}} W_{\rm{S}})\hbar P(E)$.
% \begin{equation}
%     D^{\slashed{P},\slashed{T}}E=-\frac{1}{2}(d_e W_d+C_{\rm{S}} W_{\rm{S}})P(E).
% \end{equation}
$W_d, W_{\rm{S}}$ are known as enhancement factors and $P(E) \in [0,1]$ is the polarization factor \cite{haase2021}.
The presently most sensitive experimental searches employ spin-precession methods, where a superposition of two EDM-sensitive Zeeman sublevels with energy difference $2(\mu {B}\pm D^{\slashed{P},\slashed{T}} {E})$ is employed. 
During a time $T$ in magnetic and electric fields, this superposition state rotates by an angle 
\begin{equation}\label{eq:phase}
\phi=2(\mu {B}\pm D^{\slashed{P},\slashed{T}} {E})  T/\hbar.
\end{equation}
A detailed understanding of the spin-precession process is crucial since the precession phase associated with a limit on $d_e < 10^{-30} e~\text{cm}$ ranges from mrad to nrad for searches in diatomic molecules. The experimental challenge consists in disentangling molecular effects such as the Zeeman effect, Stark effect and light shifts from the EDM contribution, where the latter part changes sign with reversal of the relative orientation of $\vec{E}$ and $\vec{B}$ fields. 
The ability to separate these effects depends on the intrinsic sensitivity of the investigated system, the statistical precision of the measurement, and limiting of systematic biases. The statistical limit on the spin-precession phase, given by the quantum projection limit $\delta\phi=\frac{1}{\sqrt{N}}$ for $N$ measurements, results in a sensitivity on new-physics parameters such as $d_e$ of
\begin{equation}\label{eq:statisticallimit}
    \delta d_e = \frac{1} {W_d P(E) T \sqrt{N} }.
\end{equation}
This shows the benefit of a large particle number $N$, long interaction time $T$ and a high enhancement factor $W_d$ and polarization $P(E)$. Systematic biases depend on the design and execution of the measurement.

In this letter, we provide a method to limit systematic errors on the interpretation of the results by determining key experimental parameters during an EDM-sensitive measurement. We have developed a description of the measurement process using the optical Bloch equations (OBE) and compare this to experimental results obtained using barium monofluoride molecules (BaF). 

\textit{Theory.}---
The superposition state, Eq.~\ref{eq:superposition}, is created and read out by an off-resonant two-photon process, where the ground state $X ^2\Sigma^+,N=0$ hyperfine levels are coupled via the $A\, ^2\Pi_{1/2}$ manifold (Fig.~\ref{fig:expdescription}). Molecules prepared in the $\ket{0,0}$ level are transferred by a laser-light pulse with two-photon Rabi frequency $\Omega_{PS}$,
% footnote{The two-photon Rabi frequency is in the limit of $\Delta\gg \Omega_P,\Omega_S$ equal to $\Omega_{PS}=-\frac{\Omega_P \Omega_S}{2\Delta}$}, 
frequency difference $\omega_{PS}=\omega_P-\omega_S$ and length $t$ to the superposition state
\begin{equation}
\ket{\psi} = \alpha \ket{1,1} + \alpha' \ket{1,-1} + \beta \ket{1,0} + \gamma \ket{0,0},
\label{eq:superposition}
\end{equation}
with the notation $\ket{F,m_F}$ for the 
$X^2\Sigma^+,N=0$. For two-photon detuning $\delta=\omega_{PS}-\omega_{\text{HFS}}(E)$ equal to zero (two-photon resonance) and $\Omega_{PS} t =\pi$, the molecules are transferred from $\ket{0,0}$ to $\ket{\psi}$ with $\alpha = \alpha'=\frac{1}{\sqrt{2}}$ and 
$\beta=\gamma=0$, which is referred to as a $\pi$-pulse. 
In general, an incomplete population transfer from $\ket{0,0}$ to $\ket{1,\pm1}$ occurs and the system is transferred to a superposition state for which $\alpha,\alpha',\beta,\gamma\neq0$. 
After an interaction with a second laser-light pulse with the same Rabi frequency $\Omega_{PS}$ and length $t$, the spin-precession phase is read out by measuring the populations $P_i$ in the $F=0$ and $F=1$ levels.
The populations after the spin-precession sequence $P_i$  \footnote{An analytical approximation for $P_i$ for a coupling by radio-frequency fields and limiting to the $\ket{1,1}, \ket{1,-1}$ and $\ket{0,0}$ can be found in \cite{kara2012measurement}.} depend not only on the precession phase $\phi$. $P_i$ is a function of the detunings $\delta$ and $\Delta$ (including Doppler shifts), the Rabi frequencies $\Omega_{P/S}$, the polarizations, $\hat{e}_{P/S}$, the phase differences between the two laser-light pulses, $\Phi_{P/S}$, the external fields $\vec{E}$ and $\vec{B}$, the lengths of the pulses $t$, the period between the start of the first and the second pulse $T$, and the initial state $\rho_0$, i.e.
\begin{equation}\label{Pgeneral}
     P_i=P_i(\delta, \Delta,\Omega_{P/S},\hat{e}_{P/S},\Phi_{P/S},\vec{E},\vec{B},t,T,\rho_0). 
\end{equation}

For two $\pi$-pulses, the experiment reduces to an effective two-level spin-precession experiment of levels $\ket{1,\pm1}$, and the population in $F=1$ after the second pulse is  
$P_{F=1}=\sin^2\frac{\phi}{2}$. In case of incomplete population transfer, coherence builds up between the superposition state and the two-photon laser field with a phase $\theta = \delta\cdot T$.

The OBE describing the dynamics of the eight levels involved,
\begin{equation}\label{eq:OBE}
    \frac{\partial \rho}{\partial t}=\frac{1}{i \hbar}[H(t), \rho]+L_{\text {relax}}(\rho),
\end{equation}
are solved to obtain the populations $P_{i}$ after the spin-precession sequence. The density matrix $\rho$ describes an eight-level system (Fig.~\ref{fig:expdescription}) and contains the level populations $P_i$ and coherences between the levels. The Hamiltonian $H(t)$ describes the energies of the eight levels, taking into account Stark and Zeeman shifts in applied external electric and magnetic fields, and couplings due to two applied laser fields. The effect of damping is included in $L_{\text{relax}}$, in particular spontaneous decay to the ground state and losses to other vibrational and rotational states. 

The OBE (Eq.~\ref{eq:OBE}) are constructed as in \cite{EADijck2015} and solved in MATLAB by extending the code from \cite{Oberst1999}. The populations $P_i$ after the spin-precession sequence are obtained as function of the parameters in Eq.~\ref{Pgeneral}. The numerical solution includes incoherent effects, light shifts, effects of all hyperfine levels and imperfect laser-light polarization.

In the NL-$e\rm{EDM}$ experiment, $t,T,\delta,\Delta,\hat{e}_{P/S}$ and $\Phi_{P/S}$ are well controlled, see experiment section. However, the external electric and magnetic fields and two-photon Rabi frequency have to be measured in the experiment. With the OBE we investigate how the measured populations $P_i$ after the spin-precession sequence depend on these parameters and how their effects can be determined from $P_i$.
The magnetic and electric field determine the phase $\phi$ due to the Zeeman effect and EDM (Eq.~\ref{eq:phase}), and the electric field affects the phase $\theta=\delta\cdot T$ by the tensor Stark effect, since $\delta=\omega_{PS}-(\omega_{\text{HFS}}^0-\omega_\text{tensor}(E))$~\cite{Vutha2018}. 
\begin{figure}[t]
    \centering
    \includegraphics[width=\linewidth]{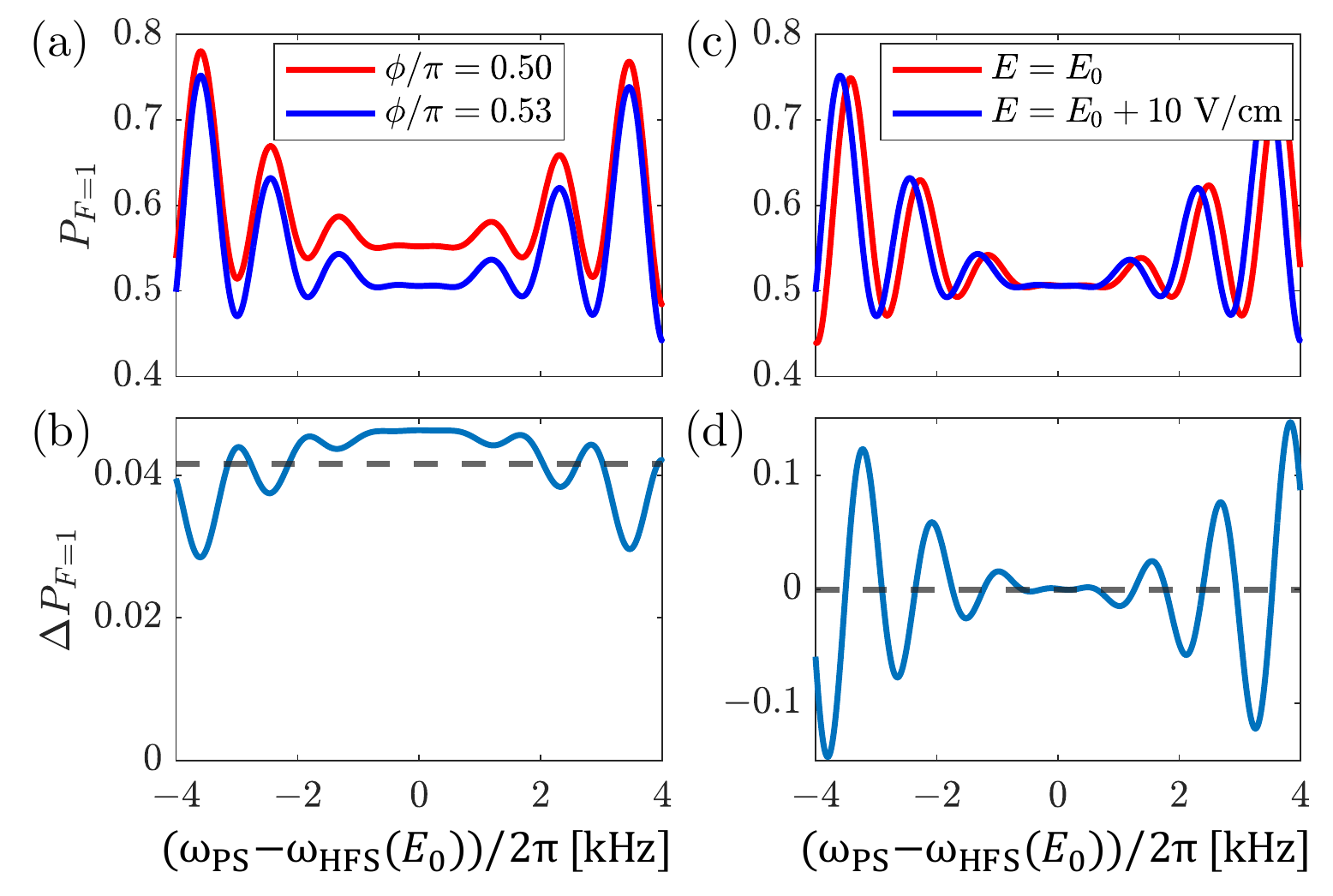}
    \caption{(a) Calculation of $P_{F=1}$ for different two-photon detuning $\delta=\omega_{PS}-\omega_{\rm{HFS}}(E)$ around the working point of the experiment, i.e. $\Omega_{PS}t\approx\pi$, and $E_0=2~\mathrm{kV/cm}$, for $\phi/\pi=0.50$ (red) and $\phi/\pi=0.53$ (blue). (b) The difference between the red and blue curves in (a), i.e. $\Delta P_{F=1}$ for a change of $\Delta\phi=0.03\pi$ as function of the detuning. At $\delta=0$, $P_{F=1}\approx \sin^2(\phi/2)$ and therefore $\Delta P_{F=1}\approx\Delta\phi/2$. The dotted line is $P_{F=1}$ averaged over the interval $\delta=-4~\text{kHz}$ to $\delta=4~\text{kHz}$. (c) $P_{F=1}$ as function of $\delta$ for $\Omega_{PS}t\approx\pi$ and $\phi=\pi/2$, for electric fields $E=E_0$ (red) and $E=E_0+10~\text{V/cm}$ (blue). A change in the electric field results in a shift of the spectrum, for an electric-field dependence of the tensor Stark shift $\omega_{\text{tensor}}$ of 14.9~kHz/(kV/cm). (d) The difference between the red and blue curves in (b), i.e. $\Delta P_{F=1}$ for a change of electric field $\Delta E=10~\text{V/cm}$. This provides the sensitivity to the externally applied electric field due to $\omega_{\text{tensor}}(E)$. The average value (dotted line) remains zero in this case. The large values of $\Delta\phi$ and $\Delta E$ are chosen for visibility.}
    \label{fig:sensitivityspectrum}
\end{figure}

The effects of the two phases $\phi$ and $\theta$, have clearly identifiable signatures which can be distinguished by analyzing $P_i$ as function of the laser-light frequency $\omega_{PS}$, i.e. the spin-precession spectrum, as shown in Fig.~\ref{fig:sensitivityspectrum}. For an increase of $\phi$, the population $P_{F=1}$ increases for every $\omega_{PS}$ (Fig.~\ref{fig:sensitivityspectrum}a), while a change in $\theta$ due to the electric field results in a shift of the spectrum (Fig.~\ref{fig:sensitivityspectrum}c). The spin-precession spectrum thus allows to distinguish a change in $P_i$ due to the phase $\phi$ from a change due to the phase $\theta$. The OBE model enables to extract $\phi$ and $\theta$ and their precision increases with the same statistics~(Eq.~\ref{eq:statisticallimit}). The dashed line in (Fig.~\ref{fig:sensitivityspectrum}b) is the average of $\Delta P_{F=1}$ over the 8~kHz region. The sensitivity to $\phi$ and hence the EDM only reduces by about 10\% w.r.t. the sensitivity at $\delta=0$.

\begin{figure}[t]
\centering
\includegraphics[width=\linewidth]{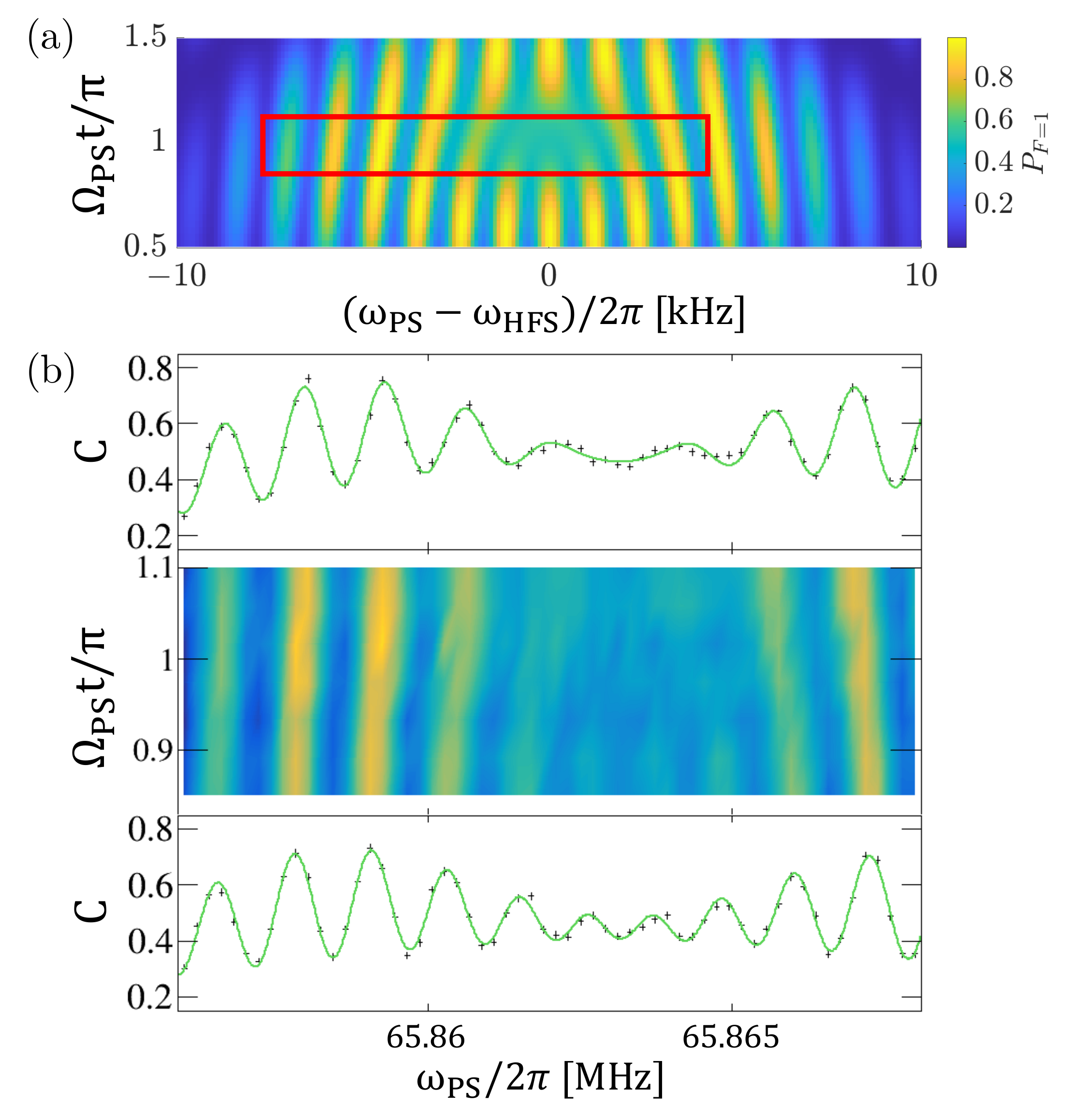}
\caption{(a) Calculation of $P_{F=1}$ as function of $\delta=\omega_{PS}-\omega_{\rm{HFS}}(E_0)$ and state rotations of $\Omega_{PS} t$ ranging from $\pi/2$ to $3\pi/2$. The magnetic field $B$ provides a phase $\phi=\pi/2$ (Eq.~\ref{eq:phase}). Note, that the number of interference fringes is different by 2 for $\Omega_{PS} t/\pi$ larger, respectively smaller than 1. The red box indicates the region measured in (b). (b) Observed fringe pattern for different $\Omega_{PS}t$, at $\phi\approx\pi/2$. The spectra at the top and the bottom are at $\Omega_{PS}t/\pi=1.092(6)$ and $\Omega_{PS}t/\pi=0.853(6)$. The line through the data points is the result from OBE. The uncertainties on the data points result from photon counting statistics.}
\label{fig:Fringes_vs_intensity_exp}
\end{figure}

The spin-precession spectrum contains information about the Rabi frequency $\Omega_{PS}$. In Fig.~\ref{fig:Fringes_vs_intensity_exp}a the population $P_{F=1}$ is calculated around the working point of the experiment. A striking feature is that when changing $\Omega_{PS} t$ from smaller than $\pi$ to larger than $\pi$, the number of fringes reduces by two. The measurement of the spin-precession spectrum is thus not only a measurement on the phase $\phi$, but also a measurement of the electric field and the Rabi frequency, and can therefore be used to limit systematic biases on the interpretation of the signal in terms of the EDM.

\begin{figure}[t]
    \centering
    \includegraphics[scale=0.375]{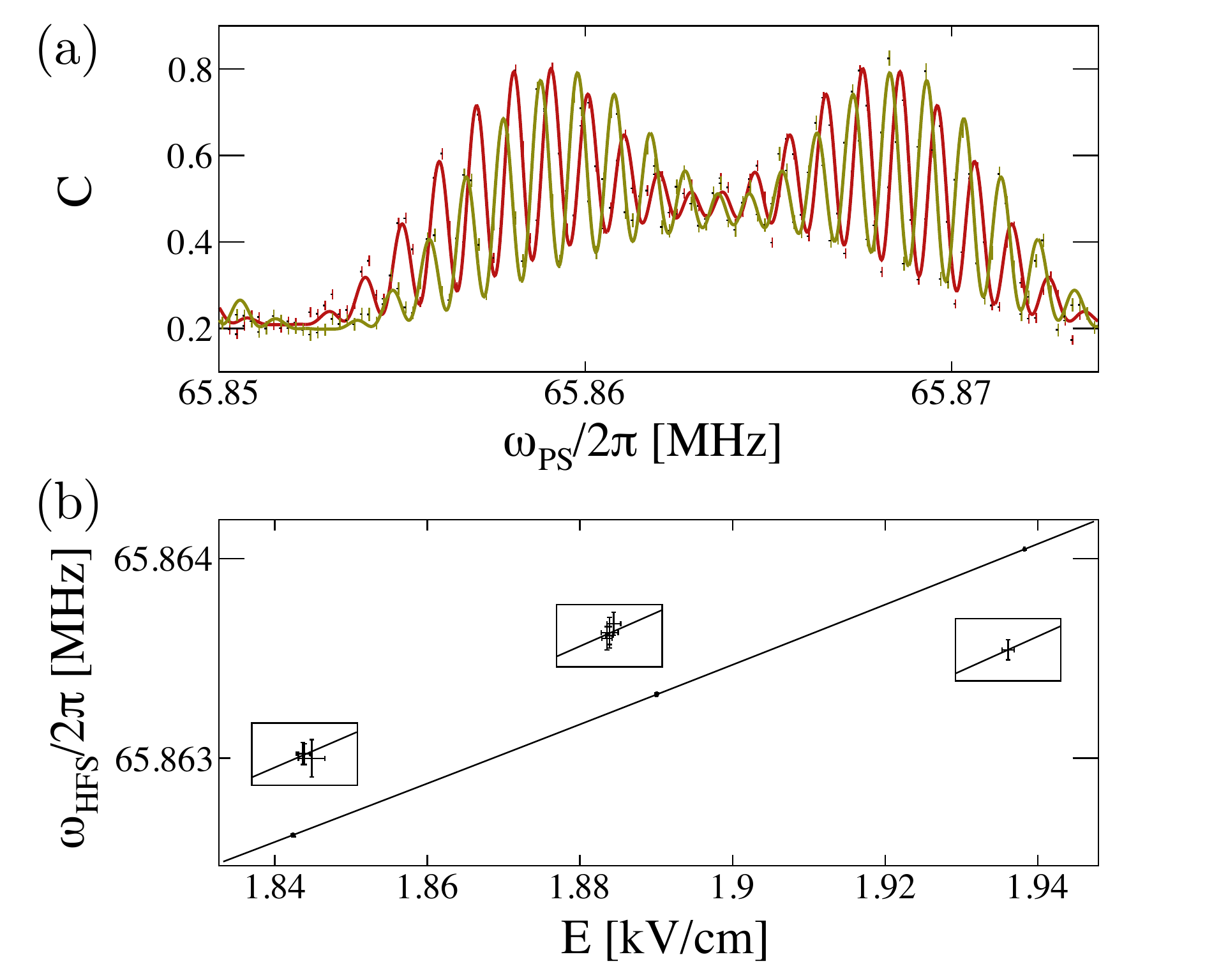}
    \caption{(a) Observed interference pattern for electric field $E=1.8900(3)$~kV/cm (green) and $E=1.9383(3)$~kV/cm (green). The electric-field-dependent hyperfine structure $\omega_{\rm{HFS}}(E)$ is determined by the center frequency of the interference pattern. The magnetic field is $B=4.04(7)$~nT and the timings are $T = 1$~ms and $t=80~\mu\rm{s}$. The contrast C is the experimental realisation of $P_{F=1}$. The uncertainties of the data points are determined by the counting statistics of the fluorescence signal. (b) The hyperfine structure splitting  $\omega_{\rm{HFS}}(E)$ changes with electric field by $14.99(7)$~Hz/(V/cm) at $E=1.9$~kV/cm. The three insets are at 100 times enlarged scale to show the uncertainties (i.e. 1~V/cm, resp. 20~Hz).}
\label{fig:SPtwoEfields}
\end{figure}

\textit{Experiment.}---
The spin precession takes place in a homogeneous magnetic and electric-field region, the interaction zone. A five-layer \textmu-metal magnetic shield provides a shielding factor $\Delta B_{\text{ext}}/\Delta B_{\text{int}} \approx 10^{6}$, with $\Delta B_{\text{ext}}$ and $\Delta B_{\text{int}}$ respectively the external and internal magnetic field change. The innermost cylinder has a length of 130 cm and a diameter of 50 cm. Coaxial to this, a cosine coil (length 100 cm, diameter 30 cm) generates a homogeneous $B$-field of several nT and a homogeneity of $\mathcal{O}(10~\text{pT})$ orthogonal to the velocity of the molecules. A cylindrical glass vacuum chamber (pressure below $10^{-7}$~mbar) contains two parallel, 4~cm separated indium tin oxide (ITO) coated glass plates (height 10~cm, length 75~cm) generating an electric field $E$ \mbox{(anti-)parallel} to the magnetic field $B$. Electric fields up to $E=5$ kV/cm can be generated with a homogeneity of $\Delta E/E <10^{-4}$.

A 10~Hz pulsed beam of $^{138}$Ba$^{19}$F molecules from a supersonic beam source~\cite{Aggarwal2021} travels through the experimental setup. The molecules are prepared in $F=0$ by optical pumping with approximately $90~\%$ efficiency before entering the interaction zone. This corresponds to $\gamma \ket{0,0} = 0.9 \ket{0,0}$. The velocity of 610(4)~m/s (velocity spread $\Delta v/v = 0.054(9)$) yields a coherence time of up to  $T\approx 1$~ms as the molecules traverse the homogeneous field region. 

The spin-precession measurement sequence is implemented solely with optical techniques: two overlapped pulsed laser beams with a frequency difference around $\omega_{\rm{HFS}}$ couple the two hyperfine states, creating and reading out a superposition via a two-photon process. In order to have minimal phase noise the laser beams are derived from a single laser by two acousto-optical modulators (AOMs) operated at frequencies $\omega_{\rm{rf}1}$ and $\omega_{\rm{rf}2}$. The pulse timings $t$, $T$, and frequencies $\omega_{\rm{rf}1}$, $\omega_{\rm{rf}2}$, are referenced against a GPS-stabilized rubidium atomic clock.
Customizable pulse patterns are generated by gated rf-signals driving the AOMs for precise control of frequency and relative intensities. The first-order output beams of the AOMs are coupled into single-mode fibers. The outputs are overlapped on a polarizing beam splitter, ensuring orthogonal polarizations $\hat{e}_P, \hat{e}_S$ of the frequency components. The combined beams are expanded to radius 20(2)~mm with a beam divergence of less than $100~\mu$rad and are sent counter-propagating to the molecular beam with an alignment of better than $300~\mu$rad.
A two-pulse sequence (Fig.~\ref{fig:expdescription}c) yields the full spin-precession signal when varying the frequency difference applied to the AOMs $\omega_{PS} = \omega_{\rm{rf}2} -\omega_{\rm{rf}1}$ and probing the population in $F=0$. The molecules are detected 3650~mm downstream from the supersonic source by laser induced fluorescence from a fiducial region of about 1~cm diameter. The fluorescence of the $X^2\Sigma ,v=0, N=0, J=1/2, \rightarrow \Pi_{3/2},J=3/2$ transition at a wavelength of 815~nm is detected by an infrared-sensitive photomultiplier tube (PMT, H7422-50 Hamamatsu) in photon counting mode. This transition is used to probe the ground state population due to the PMT's increased sensitivity at lower wavelengths, as well as the desirable absence of optical cycling from the transition. In addition, the scattered 860~nm two-photon transition light can be optically filtered out before the PMT due to the 45~nm difference between the transitions.

% \textit{Results.}---  
The sensitivity of spin-precession signal to the electric field $E$ due to the tensor Stark shift $\omega_{\text{tensor}}(E)$ (Fig. \ref{fig:sensitivityspectrum})b) has been experimentally verified as shown in Fig.\ref{fig:SPtwoEfields}. The shift of the spin-precession spectrum in frequency per electric field is the slope of the tensor Stark shift with electric field, $d\omega_{\text{tensor}}/dE$. We determined this slope to be 14.99(7)~Hz/(V/cm) and field changes smaller than 0.1~V/cm become observable.

The dependence of $P_{F=1}$ on the Rabi frequency has been experimentally verified (Fig.~\ref{fig:Fringes_vs_intensity_exp}) and compared to the results from the OBE calculations while keeping the timings at $t = 80~\mu s$ and $T = 0.8~$ms. A frequency range of 12~kHz around $\omega_{\text{HFS}}(E)$ permits the determination of $\Omega_{PS}t$ with an uncertainty of better than 1$\%$ (Fig.~\ref{fig:Fringes_vs_intensity_exp}b). The interference pattern is in excellent agreement with the OBE calculations (Fig.~\ref{fig:Fringes_vs_intensity_exp}b). The determination of these parameters improves with larger photon counting.

\textit{Conclusion and outlook.}--- 
Major systematic biases on the interpretation of spin-precession measurements in EDM searches arise from the control of experimental parameters such as the electric-field strength and spin rotation $\Omega_{PS}t$. 
We demonstrated the implementation of an all-optical spin-precession method which exploits the complex interference signal. 
The combination of laser fields counter-propagating to the molecular beam for the state manipulation and a calculation of the state evolution in an OBE framework, permits precise measurements of the experimental parameters when a frequency range around the hyperfine structure splitting $\omega_{\text{HFS}}$ is observed. This reduces significantly the amount of necessary auxiliary measurements of experimental parameters without compromising the statistics for EDM searches. This method is particularly suited for precision experiments relying on spin-precession methods in systems with more than two levels. This is being exploited in the ongoing NL-\it{e}\rm{EDM} experiment. 
% \vspace{5pt}

% \section*{Acknowledgements}
We acknowledge the contribution of R. Borchers, J. van Driel, T. E. Tiemens, I. E. Thompson and B. Schellenberg during data taking. We would like to thank L. Huismann and O. B\"oll for technical support. The NL-\it{e}\rm{EDM} consortium receives program funding (\it{e}\rm{EDM}-166) from the Netherlands Research Council (NWO).

\bibliography{BaF_total}

% \appendix
% \section{Appendix section}
% \label{Appendix_A}

\end{document}